# Note on "A. Gorguis, A reliable approach to the solution of Navier–Stokes equations, Appl. Math. Lett. 25 (2012) 2015-2017"

TAREK M. A. EL-MISTIKAWY

Department of Engineering Mathematics and Physics, Cairo University, Giza 12211, Egypt

**Abstract**

Gorguis' claim of being able to transform Navier-Stokes equations into linear ones through the Cole-Hopf transformation is disputed. It is shown that the cases treated by Gorguis are intrinsically linear; involving a velocity potential $\psi$ that is governed by Laplace's equation. They require the external force to be conservative and the initial and boundary conditions to admit such cases of fluid flow. The pressure cannot be known a priori, as suggested by Gorguis, but is determined so that it can be consistent with $\psi$. Other errors in the cited article are also indicated.

**Introduction and Analysis**

In her article [1], A. Gorguis allegedly describes a method for transforming the nonlinear axisymmetric form of the incompressible Navier-Stokes equations into linear equations, through the Cole-Hopf transformation.

The governing equations are

$$\nabla \cdot \mathbf{V} = 0 \tag{1}$$

$$\frac{\partial \mathbf{V}}{\partial t} + \mathbf{V} \cdot \nabla \mathbf{V} - \nu \Delta \mathbf{V} = -\frac{1}{\rho} \nabla p + \mathbf{f} \tag{2}$$

where $\rho$ is the density, $\nu$ is the kinematic viscosity, $t$ is the time, $p$ is the pressure, $\mathbf{V}$ is the velocity vector, and $\mathbf{f}$ is the external force. $\nabla$ and $\Delta$ are the gradient and Laplacian operators, respectively.

For the considered axisymmetric case, the velocity vector in cylindrical coordinates $(r,\theta,z)$ writes $\mathbf{V}=(u(r,z;t),0,w(r,z;t))$.

Gorguis introduces a scalar function $\psi(r,z;t)$ such that

$$u=\psi_r \text{ and } w=\psi_z \tag{3}$$

then applies the transformation

$$\psi=-2\nu\log\phi \tag{4}$$

Equations (3) mean that $\mathbf{V}=\nabla \psi$. Hence $\psi$ is, in fact, a velocity potential; and the flow is irrotational $\nabla \times \mathbf{V}=\mathbf{0}$. It is governed by the linear Laplace's equation

$$\Delta \psi = 0 \tag{5}$$

where the continuity equation (1) has been used.

Obviously, there is no need for the Cole-Hopf transformation to linearize Eq. (2).

The types of viscous flow treated by Gorguis are, in effect, inviscid; diffusion of information through viscosity being inactive. This requires the external force $\mathbf{f}$ to be conservative of potential $T(r,z,t)$, say; so that $\mathbf{f}= (-T_r, 0, -T_z)$.

Introduction of $\psi$ into the components of Eq. (2) leads to (Compare with Eqs. (11) and (12) of [1].)

$$(\psi_r)_t + \psi_r(\psi_r)_r + \psi_z(\psi_r)_z - \nu\{(\psi_r)_{rr} + \frac{1}{r}(\psi_r)_r - \frac{\psi_r}{r^2} + (\psi_r)_{zz}\} = -\frac{1}{\rho}p_r - T_r \tag{6}$$

$$(\psi_z)_t + \psi_r(\psi_z)_r + \psi_z(\psi_z)_z - \nu\{(\psi_z)_{rr} + \frac{1}{r}(\psi_z)_r + (\psi_z)_{zz}\} = -\frac{1}{\rho}p_z - T_z \tag{7}$$

Integration of Eqs. (6) and (7) with respect to $r$ and $z$, respectively, gives (Compare with Eqs. (13) and (14) of [1], in which the right hand sides were not integrated.)

$$\psi_t + \tfrac{1}{2}(\psi_r)^2 + \tfrac{1}{2}(\psi_z)^2 - \nu\{\psi_{rr} + \frac{1}{r}\psi_r + \psi_{zz}\} = -\frac{p}{\rho} - T + C_{,r} \tag{8}$$

$$\psi_t + \tfrac{1}{2}(\psi_r)^2 + \tfrac{1}{2}(\psi_z)^2 - \nu\{\psi_{rr} + \frac{1}{r}\psi_r + \psi_{zz}\} = -\frac{p}{\rho} - T + C_{,z} \tag{9}$$

where $C_{,r}(z,t)$ and $C_{,z}(r,t)$ are arbitrary functions of integration. Noting that the left hand sides of Eqs. (8) and (9) are identical, the initial and boundary conditions must give $C_{,r}(z,t)=C_{,z}(r,t)=C(t)$ for the right hand sides to be identical, for consistency. $C(t)$ may, then, be absorbed in $T$, leading to

$$\psi_t + \tfrac{1}{2}(\psi_r)^2 + \tfrac{1}{2}(\psi_z)^2 - \nu\{\psi_{rr} + \frac{1}{r}\psi_r + \psi_{zz}\} = -\frac{p}{\rho} - T \tag{10}$$

Thus, having solved the linear Laplace equation (5) for $\psi$, Eq. (10) is used to get the corresponding pressure $p$, given $T$.

Equation (10) can be rewritten using the transformation (4). This leads to (Compare with Eq. (18) of [1], in which $\Delta\phi$ was erroneously given by $\phi_{rr}+\phi_{zz}$.)

$$\phi_t - \nu\{\phi_{rr} + \frac{\phi_r}{r} + \phi_{zz}\} = \frac{\phi}{2\nu}(\frac{p}{\rho} + T) \qquad (11)$$

It is noted, further, that applying the transformations (3) and (4) to Eq. (6) of [1] does not lead to Eq. (20) of [1]. What does, is applying the transformations to the physically unjustified equation $u+w=0$.

**Conclusion**

Gorguis' claim of being able to transform Navier-Stokes equations into linear ones through the Cole-Hopf transformation has been shown to be invalid.